\documentclass[11pt,letterpaper]{article}

\usepackage{amsmath}
\usepackage{amsfonts}
\usepackage{amssymb}
\usepackage{amsxtra}
\usepackage{comment}
\usepackage{epsfig}
\usepackage{graphicx}
\usepackage{enumerate}
\usepackage{float}
\usepackage{indentfirst}
\usepackage{graphicx}
\usepackage{verbatim}
\usepackage{setspace}
\usepackage{color}

\newtheorem{assumption}{\textbf{Assumption}}
\newtheorem{problem}{\textbf{Problem}}

\newtheorem{theorem}{\textbf{Theorem}}

\newtheorem{fact}{\textbf{Fact}}

\newcommand{\qed}{\begin{flushright}$\Box\Box\Box$\end{flushright}}

\newcommand{\lpt}{\left.}
\newcommand{\rpt}{\right.}
\newcommand{\lc}{\left[}
\newcommand{\ly}{\left\{}
\newcommand{\ry}{\right\}}
\newcommand{\rc}{\right]}
\newcommand{\lp}{\left(}
\newcommand{\rp}{\right)}

\newcommand{\bs}[1]{\boldsymbol{#1}}

\title{\bf Optimal metabolic pathway activation\footnote{An updated version of this manuscript will be presented at the 17th IFAC World Congress, Seoul, Korea, 2008}}

\author{D. Oyarz\'un\thanks{Corresponding author: diego.oyarzun@nuim.ie}, B. Ingalls, R. Middleton and D. Kalamatianos\footnote{D. Oyarz\'un, R. Middleton and D. Kalamatianos are with the Hamilton Institute, National University of Ireland, Maynooth, Ireland. B. Ingalls is with the Department of Applied Mathematics, University of Waterloo, Canada.}}

\usepackage{amsmath}
\usepackage{amssymb}

\begin{document}
\maketitle

\begin{abstract}
This paper deals with temporal enzyme distribution in the activation of biochemical pathways. Pathway activation arises when production of a certain biomolecule is required due to changing environmental conditions. Under the premise that biological systems have been optimized through evolutionary processes, a biologically meaningful optimal control problem is posed. In this setup, the enzyme concentrations are assumed to be time dependent and constrained by a limited overall enzyme production capacity, while the optimization criterion accounts for both time and resource usage.
Using geometric arguments we establish the bang-bang nature of the solution and reveal that each reaction must be sequentially activated in the same order as they appear in the pathway. The results hold for a broad range of enzyme dynamics which includes, but is not limited to, Mass Action, Michaelis-Menten and Hill Equation kinetics. 
\end{abstract}

\section{Introduction}

Metabolic networks consist of pathways of biochemical reactions which consume and produce required biomolecules.  Pathway performance depends on network structure and the kinetics of the enzymes which catalyze each interaction \cite{heisch96}, leading to diverse dynamic behaviors in terms of stability, steady state and transient response.  Most cellular processes rely on the appropriate operation of some set of pathways and so their behaviour underpins functional requirements for cellular operation. It has been proposed that metabolic dynamics play a significant role in cell fitness, and so metabolic system design has been optimized through evolutionary processes \cite{hescho91}.

In this paper we address the mechanism responsible for the distribution of enzyme concentrations in a metabolic pathway in such a way that a meaningful optimality criterion is satisfied. Previous studies have tackled this problem by considering a number of objective functions, e.g., flux optimization \cite{hescho91,heikli96,holzhu04}, minimization of total enzyme concentration \cite{klihei99} and maximization of growth rate \cite{bilu_etal06}. All these works focus on the steady state properties of the pathway and limit the analysis by considering constant enzyme concentrations \cite{hescho91,heikli96,holzhu04,klihei99,bilu_etal06}. Nonetheless, the temporal distribution of the enzyme concentrations may have a critical impact on the pathway behavior. Well defined hierarchical temporal patterns have been observed in enzyme expression levels in amino acid \cite{zaslaver_etal04} and \emph{E. coli} flagella biosynthesis \cite{kalir_etal01}. In addition,  certain metabolic pathways are naturally at rest and become active under changes in external conditions or when certain key biomolecules are required. In cases where the cellular response impacts fitness, we may presume that the activation of these pathways will occur as quickly as possible.  Since the cell has a limited biosynthetic capacity there must be a mechanism that accounts for appropriate enzyme allocation over the activation period.

Time dependent optimal enzyme profiles have been considered in recent papers \cite{klheho02,zaslaver_etal04,oyinka07c} considering the case of unbranched pathways. Though using different approaches, all these works conclude that sequential behaviors, such as the one experimentally shown in \cite{zaslaver_etal04}, can be rationalized by optimality principles. The main drawback behind these results lies in their lack of generality, either because only a particular case is solved \cite{zaslaver_etal04}, or because simple (and often unrealistic) enzyme kinetics are assumed \cite{klheho02,oyinka07c}.

As in \cite{oyinka07c}, the motivation for this work is the observation that the problem dealt with in \cite{klheho02} can be naturally posed in terms of classical optimal control \cite{ponteal62}. Here we formulate a similar problem in a standard control theoretic framework and, using a cost functional that accounts for both time and enzyme usage, we extend the results of \cite{oyinka07c} in three directions: (a) the developed framework allows mass exchange between the pathway and its environment; (b) the results show that the reactions must be fully activated one at a time, following the same sequence as they appear in the pathway; (c) the derived qualitative behavior is valid for a broad class of enzyme kinetics that includes, but is not restricted to, the common Mass Action, Michaelis-Menten and Hill Equation kinetics \cite{cornish-bowden04}.

The main theoretical tool used throughout this paper is Pontryagin's Minimum Principle (PMP) \cite{ponteal62}. Together with simple geometric arguments, the PMP provides qualitative information regarding the bang-bang form of the optimal solution and the sequence in which the activations must be performed. We emphasize that in this paper we do not attempt to obtain a complete solution of the optimization problem, but rather we aim at deriving qualitative insights into the solution that extend and are consistent with the previous attempts to explain the sequential behavior seen in \cite{zaslaver_etal04}. Similarly, it must be stressed that the approach taken in this paper should not be confused with a time-optimal control \emph{design} problem, since our idea is to rigourously \emph{explain} a behavior that is already present in some biological systems. Besides the derivation of the main results, this paper intends to set a bridge between well known results pertaining to control theory and the systems theoretical analysis of biological systems, an approach that is recently emerging as a cornerstone in the systems biology field.

\section{Preliminaries}
In this section, we briefly state the main results of optimal
control theory developed by Pontryagin and co-workers \cite{ponteal62}. We are interested in dynamical systems of the form
\begin{align}
  \dot{x}(t)&=f\lp x(t),u(t)\rp,\label{eq:stateeq}\\
  x(0)&\in\mathbb{R}^n,
\end{align}
where $x(t)\in\mathbb{R}^n$ is the state vector, $u(t)\in\mathbb{R}^p$ is the control input vector, and
$f\lp x(t),u(t)\rp$ is a continuously differentiable function. Suppose
that the control objective is to drive the state $x(t)$ from $x(0)$
to a final condition $x(t_f)\in \mathcal{S}$, where $\mathcal{S}\subseteq \mathbb{R}^n$. We denote the set of admissible values for $u(t)$ as
$\mathcal{U}$, 
and let $\mathcal{M}_{\mathcal {U}}$ be the set of piece-wise continuous functions $u: [0, t_f) \rightarrow \mathcal{U}$.
The objective is
then to determine an optimal control input, 
$u^*(\cdot)$, such that
\begin{align}
u^*(\cdot)&=\arg\min_{u(\cdot)\in\mathcal{M}_{\mathcal{U}}} \mathcal{J},
\end{align}
where the target cost functional is
\begin{align}
  \mathcal{J}&=q\lp x(t_f)\rp + \int_0^{t_f} g\lp x(t),u(t)\rp
  dt.\label{eq:defJgral}
\end{align}
Define the Hamiltonian as
\begin{align}
  \mathcal{H}\lp x(t),u(t),p(t)\rp&=g\lp x(t),u(t)\rp + p(t)^T f\lp x(t),u(t)\rp,\label{eq:defHam}
\end{align}
where the vector $p(t)\in\mathbb{R}^n$ is the system's co-state. The PMP gives the following set of necessary conditions for optimality: if an optimal $u^*(\cdot)$ exists, then there exist nontrivial trajectories $x^*(\cdot)$ and $ p^*(\cdot)$ such that:
\begin{enumerate}[(a)]
  \item  The set of differential equations
\begin{align}
  \dot{x}^*(t)&= \frac{\partial \mathcal{H}\lp x^*(t),u^*(t),p^*(t)\rp}{\partial p},\label{eq:pont1}\\
  \dot{p}^*(t)&= -\frac{\partial \mathcal{H}\lp x^*(t),u^*(t),p^*(t)\rp}{\partial x},\label{eq:pont2}
\end{align}
is satisfied subject to the boundary conditions $x^*(0)=x(0)$ and $x^*(t_f)\in \mathcal{S}$.

  \item The Hamiltonian is minimized by the optimal control input
  $u^*(t)$ for all $t\in\lc 0,t_f\rc$, i.e.,
  \begin{align}
    \mathcal{H}\lp x^*(t),u^*(t),p^*(t)\rp &=
    \min_{u(t)\in\mathcal{U}} \mathcal{H}\lp x^*(t),u(t),p^*(t)\rp.\label{eq:pont3}
  \end{align}

  \item The Hamiltonian for the optimal control input is zero for
  all $t\in\lc 0,t_f\rc$, that is,
  \begin{align}
    \mathcal{H}\lp x^*(t),u^*(t),p^*(t)\rp&=0,\,\forall\, t\in\lc 0,\,t_f\rc\label{eq:pont4}
  \end{align}

\item The costate satisfies the terminal condition $p^*(t_f) \in \mathcal{S}^\bot$, where $\mathcal{S}^\bot$ is orthogonal to $\mathcal{S}$ at $x(t_f)$.

\end{enumerate}

\section{Problem formulation}
\subsection{General setup}
\begin{figure}[H]
\centering
\input{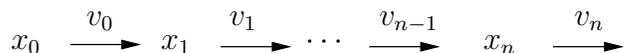}
\caption{Unbranched metabolic pathway.}\label{fig:unbranched}
\end{figure}

We deal with unbranched metabolic pathways as depicted in Fig. \ref{fig:unbranched}, where $x_0$ denotes the concentration of substrate feeding the pathway, $x_i$ is the concentration of the $i$th intermediate metabolite and $v_i$ is the chemical rate characterizing the $i$th reaction. 
In general, we assume that $v_i(t)=v_i\lp x_i(t),u_i(t)\rp$, where $u_i(t)$ is the concentration of the enzyme catalyzing the $i$th reaction.

Following established techniques for analyzing metabolic pathways, the substrate $x_0$ is assumed to change in a time scale that is considerably slower than the pathway in which we are interested and therefore it is considered as a constant in our analysis.

Each rate law $v_i$ characterizes the kinetic properties of one of the enzymes catalyzing the reactions in the pathway.  These are typically nonlinear in the metabolite concentration $x_i$, so a general analysis is often not tractable.  However the results presented in this paper hold for a fairly broad class of enzyme kinetics, namely those satisfying the following assumptions.

\begin{assumption}\label{ass:ratelaws}$\,$
\begin{enumerate}
\item[1A] Each rate law is linear in the enzyme concentration, i.e., they can be written as
    \begin{align}
    v_i(x_i(t),u_i(t))&= w_i(x_i(t))u_i(t),\,\forall\, i=0,1,\ldots,n, \label{eq:ass1:vlinearinu}
    \end{align}
    where $w_i(x_{i}(t))$ does not depend on $u_i(t)$.

\item[1B] It holds
    \begin{align}
    w_i(0)&=0,\,\forall\, i=0,1,\ldots,n, \label{eq:ass1:wnull}\\
    \frac{d w_i}{d x_i}&>0,\,\forall\, x_i \in \lp 0, \infty \rp,\,\forall\, i=0,1,\ldots,n,\label{eq:ass1:partialw}
    \end{align}
\end{enumerate}
\end{assumption}
Assumption \ref{ass:ratelaws}A is satisfied by most enzyme kinetic models \cite{cornish-bowden04}, while \eqref{eq:ass1:wnull} in Assumption \ref{ass:ratelaws}B is trivially satisfied since a nonzero concentration $x_i(t)$ is required for reaction $i$ to occur. Equation \eqref{eq:ass1:partialw} states no more that, for a constant enzyme concentration, an increase in substrate $x_i$ yields an increase in the rate, possibly reaching a saturation state when $x_i$ grows unbounded. Therefore, Assumption \ref{ass:ratelaws} is satisfied by a broad class of enzyme dynamics which includes, in particular, Mass Action, Michaelis-Menten, and Hill Equation kinetics. Unless otherwise stated, for the sake of notational simplicity in what follows we will only write time $t$ as independent variable, so that expressions like $v_i(x_i(t),u_i(t))$ will be denoted just as $v_i(t)$.

The ODE model for the pathway in Fig. \ref{fig:unbranched} is given by conservation of mass as
\begin{align}
\dot{x}_i(t)&= v_{i-1}(t) - v_i(t),\,\forall\, i=1,2,\ldots,n,\label{eq:gralODE}
\end{align}

To account for the positivity of the enzyme concentrations and the limited total enzyme availability, we will impose the constraint $u(t)\in\mathcal{U}$, where $\mathcal{U}$ is a simplex in $\mathbb{R}^{n+1}$ given by
\begin{align}
    \mathcal{U}:\quad
    \ly\begin{array}{ll}
     \sum_{i=0}^{n} u_i \leq E_T\\
     u_i\geq 0,\,\forall i=0,1,\ldots,n.
    \end{array}\rpt\label{eq:defsetU}
\end{align}
For future reference we define the state, control, and flux vectors as
\begin{align}
    x(t)&= \begin{bmatrix}
    x_1(t) & x_2(t) & \cdots & x_n(t)
    \end{bmatrix}^T,\\
    u(t)&=\begin{bmatrix}
    u_0(t) & u_1(t) & \cdots & u_n(t)
    \end{bmatrix}^T,\\
    v(t)&=\begin{bmatrix}
    v_0(t) & v_1(t) & \cdots & v_n(t)
    \end{bmatrix}^T.
\end{align}

\subsection{Optimal control problem}

Assuming that the pathway is initially at rest, i.e. $x_i(0)=0,\,\forall i=1,2,\ldots,n$, we aim at obtaining enzyme temporal profiles that in the time interval $\lpt\lc 0,\,t_f\rpt\rp$ drive the pathway to a steady state characterized by a prescribed constant flux $V\in\mathbb{R}_+$. 
From Fig. \ref{fig:unbranched} and \eqref{eq:gralODE} this implies
\begin{align}
v_i(t)&= V,\,\forall\, t\geq t_f,\,\forall\, i=0,\,1,\,\ldots,\,n.\label{eq:sscondition}
\end{align}

As is clear from \eqref{eq:ass1:partialw}, each $v_i$ can (and usually does) saturate to some upper bound, so in the sequel we will assume that the target final flux $V$ is consistent with those upper bounds. The enzymes comprised in $u(t)$ should minimize a measure of the time and resource usage during the pathway activation. To that end we define the cost functional
\begin{align}
    \mathcal{J}&= \int_0^{t_f} \lp 1 + \alpha^T u(t)\rp\,dt, \label{eq:defJ}
\end{align}

where the weight $\alpha\in\mathbb{R}^{n+1}$ is entry wise nonnegative. The positivity of $u(t)$ implies that the minimization of $\mathcal{J}$ implies a combined optimization of: (i) the time taken to reach the new steady state and, (ii) a measure of the energy demanded by the enzyme usage. Regardless the result given by the optimization for $t\in\lpt \lc 0,\, t_f\rpt\rp$, we must guarantee that the steady state flux $V$ is achieved for $t\geq t_f$, which from \eqref{eq:ass1:vlinearinu} and \eqref{eq:sscondition} implies that
\begin{align}
    u_i(t) &= \frac{V}{w_i(t_f^{-})},\,\forall t\geq t_f,\,\forall i=0,1,\ldots,n.\label{eq:condonutfplus}
\end{align}

Note that \eqref{eq:condonutfplus} specifies the constant enzyme levels needed after $t=t_f$ in order to reach the target steady state flux $V$. It should be stressed that the optimization is performed over the open interval $\lpt\lc 0,\, t_f \rpt \rp$ so that the enzyme levels in \eqref{eq:condonutfplus} are not significant to the optimization. This also implies that the terminal state for the optimization problem is defined by $x(t_f^{-})=\lim_{\delta\to 0} x(t_f-\delta),\, \delta>0$. The continuity of $x(t)$ implies that the value of $x(t_f^{-})$ must be consistent with the prescribed steady state flux $V$ and the input constraints defined in \eqref{eq:defsetU}, which by using \eqref{eq:condonutfplus} implies that the final state must belong to the set
\begin{align}
    \mathcal{S}:\qquad \sum_{i=0}^{n} \frac{V}{w_i(x_{i}(t_f^{-}))} & \leq E_T,\label{eq:finalsurf}
\end{align}

In summary, the optimal control problem can be stated as follows.

\begin{problem}\label{prob:main}
Given the system \eqref{eq:gralODE} with the initial and terminal conditions
\begin{align}
    x(0)&=0,\\
    \lim_{\delta\to 0} x(t_f-\delta)&\in\mathcal{S},\quad \delta>0,
\end{align}
where $\mathcal{S}$ is the set \eqref{eq:finalsurf} and $V\in \mathbb{R}_{+}$ is a feasible flux, find
\begin{align}
u(\cdot)^* &= \arg \min_{u(\cdot)\in\mathcal{M}_{\mathcal{U}}} \mathcal{J}, \label{eq:optprob}
\end{align}
where $\mathcal{U}$ and $\mathcal{J}$ are defined in \eqref{eq:defsetU} and \eqref{eq:defJ} respectively.
\end{problem}

\section{Properties of the optimal solution}

Solving the boundary value problem (BVP) in \eqref{eq:pont1} and \eqref{eq:pont2} can be a difficult task. 
In this section we derive some properties of the optimal solution without solving the BVP, but rather by using geometric arguments together with \eqref{eq:pont3} and \eqref{eq:pont4}. The key elements in our derivations are the linearity of the Hamiltonian in the control variables and the geometry of the feasible region $\mathcal{U}$.

Referring to \eqref{eq:defHam}, \eqref{eq:ass1:vlinearinu}, and \eqref{eq:gralODE}, the Hamiltonian is given by
\begin{align}
	&\mathcal{H}\lp x(t),u(t),p(t)\rp =\nonumber\\ 
& 1 + \alpha^T u(t) + \sum_{i=1}^{n} p_i(t)\lp w_{i-1}(t)u_{i-1}(t) - w_{i}u_{i}(t) \rp,\label{eq:defHam1}\\
&\qquad\qquad\qquad\quad\,\,\,\,= 1 + \sum_{i=0}^n h_i(t) u_i(t),\label{eq:defHam2}
\end{align}
where the function $h_i(t)$ is called the $i$th switching function and, in general, depends on the concentration vector $x(t)$, the substrate $x_0$ and the co-state vector $p(t)$. The main result of this paper is presented next.

\begin{theorem}\label{thm:main}
The solution to Problem \ref{prob:main} is unique and given by 
\begin{align}
u_i^*(t)&= \ly\begin{array}{ll}
        E_T,\,&\forall\, t\in T_i\\
        0,\,&\forall\, t\notin T_i
           \end{array}\rpt,\label{eq:thm:main}
\end{align}
where $\ly T_0,\,T_1,\, \ldots,\, T_{n-1}\ry$ is a partition of the interval $\lpt\lc 0,\, t_f \rpt\rp$ such that $T_0=\lpt\lc 0,\, t_0 \rpt\rp$ and $T_i=\lpt\lc t_{i-1},\, t_i \rpt\rp,\,\forall\,i=1,\,2,\,\ldots,\,n$, with $t_i < t_j,\,\forall\, i<j$ and $t_{n-1}=t_f$.
\end{theorem}

The proof appears in the Appendix.

The result of Theorem \ref{thm:main} reflects that considerable information regarding the solution to Problem \ref{prob:main} can be obtained without the need of solving the BVP arising from applying the PMP. The metabolic pathways under consideration can exhibit nonlinear dynamics, leading to a quite involved form of the BVP and hence considerably complicating its solution. Thus, the information provided by Theorem \ref{thm:main} is a convenient way to overcome the difficulties introduced by the BVP and still be able to identify its qualitive behavior.

Moreover, this characterization of the solution makes the computation of the optimal input much easier, since one needs only to optimize over the $n$ switching times $\ly t_0,\,t_1,\,\ldots,\,t_{n-1}\ry$, rather than over the whole class of admissible inputs.

\begin{figure}[H]
  \centering
  \includegraphics[width=0.5\textwidth]{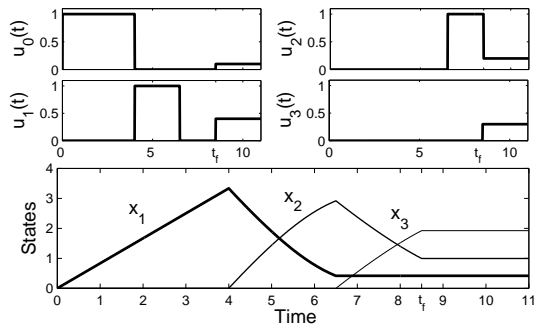}\label{fig:example}
\caption{Qualitative sketch of optimal solution for $n=3$, $E_T=1$ and $t_f=8.5$.}
\end{figure}

Equation \eqref{eq:thm:main} states that the optimal enzyme profiles are of bang-bang type and are composed of switching sequences between $0$ and the maximum level $E_T$. This clearly resembles the nature of solutions in classical time-optimal control \cite{ponteal62} and has interesting implications in the context of metabolic pathway activation. In fact, from \eqref{eq:thm:main} we first note that optimizing the pathway activation implies that for any time instant, all the available enzyme must be used, thus demanding full enzyme biosynthetic activity. Moreover, the optimization enforces activation of only one reaction at a time and most importantly, the sequence of activations follows a well defined temporal pattern. The optimal activation sequence is $\ly v_0,\, v_1,\, \ldots,\, v_{n-1}\ry$, a pattern that exactly matches the ordering in which the reactions appear in the pathway. This phenomenon has been described in \cite{klheho02, zaslaver_etal04} for the  special case of specific pathway lengths and 
reaction kinetics (Mass Action in the first case and Michelis-Menten with transcriptional regulation in the latter). Since the requirements imposed by Assumption \ref{ass:ratelaws} hold for a broad class of enzyme kinetics, Theorem \ref{thm:main} extends the finding of sequential behavior to a more general class of metabolic pathways than previously considered.
In addition, we note that reaction $v_n$ does not appear in the optimal sequence, since its activation is not required to drive the pathway to the final state, but only to achieve the target chemical flux $V$ after the time interval where the optimization is performed. The sketch in Fig.~\ref{fig:example} indicates shows a qualitative plot of the optimal solutions (with $n=3$, $E_T=1$ and $t_f=8.5$). It should be noticed that for $t\geq t_f$ the optimal enzyme profiles are obtained directly from \eqref{eq:condonutfplus}.

The generality of the results suggests that the aforementioned sequential behavior is a feature underlying time-optimal activation of unbranched pathways rather than a property arising from studying specific case. The activation sequence depends essentially on the time courses of the switching functions and, as can be checked from \eqref{eq:proofthm:hjmin} in Appendix \ref{app:proof}, the $i$th reaction is activated whenever the corresponding switching function $h_i(t)$ is the minimum of the set $\ly h_0(t),\,h_1(t),\,\ldots,\, h_n(t)\ry$. Further consideration reveals that the sequential behavior is a consequence of both pathway structure and a common property shared by the enzyme kinetics. From an intuitive point of view, the ``pipeline'' structure of the pathway implies the $i$th metabolite cannot be produced unless the upstream portion of the pathway has already been built up. Moreover, the monotonicity of the chemical rates, as expressed by \eqref{eq:ass1:partialw} in Assumption \ref{ass:ratelaws}B, precludes the possibility of having to activate an upstream reaction after the $i$th has already been activated, a fact that arises from \eqref {eq:proofseq:hellm1dot2} and \eqref{eq:proofseq:fac:1c}. This analysis suggests that the described sequential behavior might be indeed a property underpinning time-optimal unbranched pathway activation. 

The form of the optimal enzyme profiles reveals an inherent regulatory mechanism that allocates enzyme concentrations to achieve optimality. In cellular systems this regulation is performed at a transcriptional level through the regulation of the transcription factors that trigger the expression of the required enzymes. However, so far we have excluded from our analysis any sort of metabolite-level regulatory mechanisms, which in some cases are crucial to perform biological functions. This issue is briefly dealt with in the next section for a particular case.

\section{About feedback regulated pathways}

\begin{figure}[H]
    \centering
    \input{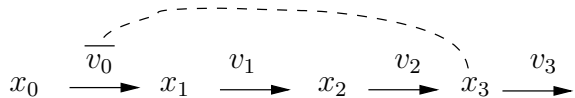}\label{fig:feedback1}
    \caption{Pathway with an allosteric feedback loop.}
\end{figure}

On the metabolic scale, regulation is often implemented through allosteric enzymes, which display a distinctive kinetic behavior \cite{cornish-bowden04}. The key idea is that the catalytic activity of an allosteric enzyme is dependent not only on its substrate, 
but also on another metabolite in the pathway. To illustrate this situation, let us consider an unbranched pathway as the one depicted in Fig. \ref{fig:feedback1}, which has an inhibitory feedback loop from $x_3$ to reaction $v_0$.
In this case we extend the definition of the reaction rate $v_0$ so that that $v_0(t)=v_0\lp x_0(t),x_3(t),u_0(t)\rp$.

A widely accepted model for allosteric kinetics \cite{monod_etal65} is given by
\begin{align}
v_0 &= \frac{k x_0/K_s \lp 1 + x_0/K_s\rp^{n-1}}{\lp 1 + x_0/K_s\rp ^n + L\lp 1+ x_3/K_I\rp^n} u_0,
\end{align}

where the parameters $\lp k,\, n,\, L,\, K_s,\, K_I\rp$ are characteristic for each allosteric enzyme.

It can be seen that, though $v_0$ is nonlinear in both $x_0$ and $x_3$, it is still linear in $u_0$. 
A close look at the proof presented in Appendix \ref{app:proof} reveals that in this case the analysis remains unchanged and therefore, the results presented in the previous sections still hold in this more complex scenario. From a control theoretic viewpoint, this is an interesting result since it states that, in this particular case, the form of the optimal solution is invariant under allosteric feedback of this kind.

On the other hand, since the feedback modifies the steady state concentrations, from a biological viewpoint this finding suggests that this sort of feedback inhibition may have a much more significant role in regulating steady state behavior than in regulating the dynamic behavior of pathway activation.  This implication has not been corroborated experimentally.

\section{Concluding remarks}

Using geometric properties of the cost functional and the feasible set of control inputs, we have derived qualitative properties behind the optimal activation of metabolic pathways. The main strength of the results lies on the fact that mild assumptions on the reaction kinetics are required, thus encompassing a broad family of monomolecular kinetics. Interesting extensions would be to include enzyme production dynamics, transcriptional regulation and the study of multimolecular reactions.

\bibliography{bibliografia}

\appendix
\section{Proof of Theorem \ref{thm:main}}\label{app:proof}

Denote the set of vertexes of $\mathcal{U}$ as $\mathcal{V}=\ly e_0,e_1,\ldots,e_n\ry \cup \ly\bs{0}\ry$, where $e_i$ has $E_T$ in its $(i+1)$th entry and $0$ elsewhere. Similarly, the set of $n-$dimension faces of $\mathcal{U}$ is defined as $\mathcal{F}=\ly F_0,\,F_1,\,\ldots,\,F_n\ry \cup \ly \mathcal{P} \ry$, where $F_i$ and $\mathcal{P}$ are the faces defined by the hyperplanes $F_i=\ly u(t)\in\mathcal{U}: u_i(t)=0\ry$ and $\mathcal{P}=\ly u(t)\in\mathcal{U}: \sum_{i=0}^{n} u_i(t)=E_T\ry$, respectively. We notice from \eqref{eq:defHam2} that $\mathcal{H}\lp x(t),u(t),p(t)\rp$ is a linear function defined over the convex polyhedron $\mathcal{U}$. Therefore, from \eqref{eq:pont3} it follows that the optimal control is located in the boundary of $\mathcal{U}$ for all $t\in\lpt\lc 0,\, t_f\rpt \rp$ and, moreover, it holds that $u^*(t)\in\mathcal{V},\,\forall t\in \lpt\lc 0,\, t_f\rpt\rp$. Let $u^{*1}(t)$ be the optimal solution for $t\in\lc t_a,\,t_b\rc$ such that $u^{*1}(t)$ is located at vertex $e_i$. If there exists another optimal solution $u^{*2}(t)\neq u^{*1}(t)$ for $t\in\lc t_a,t_b\rc$ such that $u^{*2}(t)\in F_i\setminus \mathcal{V}$, then the linearity of $\mathcal{H}$ implies that any point in $F_i$ is also optimal for $t\in\lc t_a,t_b\rc$. In particular, since $\bs{0}\in F_i, \forall i$, this implies that the origin is also optimal. However, from \eqref{eq:pont4} it holds that $\mathcal{H}$ must vanish along the optimal trajectory, which together with \eqref{eq:defHam2} makes clear that $u^*(t) \neq \bs{0},\,\forall t\in \lpt\lc  0,\, t_f\rpt\rp$, thus contradicting the optimality of the origin. Hence it follows that $u^*(t)\notin F_i\setminus \mathcal{V},\,\forall i$, so that $u^*(t)\in\mathcal{P}$ and therefore the optimal solution satisfies
\begin{align}
\sum_{i=0}^n u_i^*(t) &= E_T. \label{eq:proofthm:sumET}
\end{align}

Let $\ly T_0,\,T_1,\, \ldots,\, T_q\ry$ be a partition of the interval $\lpt\lc 0,\, t_f \rpt\rp$ such that $T_0=\lpt\lc 0,\, t_0 \rpt\rp$ and $T_i=\lpt\lc t_{i-1},\, t_i \rpt\rp,\,\forall\,i=1,\,2,\,\ldots,\,q$, with $t_i < t_j,\,\forall\, i<j$ and $t_q=t_f$. Let $U_\ell^*=\ly u_i \ry,\, i\in\mathbb{I}_\ell$, be the set of enzymes in the optimal solution that are active in the interval $T_\ell$, that is, the optimal control $u^*(t)$ lies in the convex hull of vertexes $\ly e_j\ry,\, j\in \mathbb{I_\ell},\,\forall\,t\in T_\ell$. This implies that each control satisfies $u_i(t)> 0,\,\forall\,t\in T_\ell,\,\forall\,i\in\mathbb{I}_\ell$ and they belong to the surface
\begin{align}
\sum_{i\in\mathbb{I}_\ell} u_i(t) &= E_T.\label{eq:proofthm:surfuj}
\end{align}

We consider the partition $\ly T_0,\,T_1,\, \ldots,\, T_q\ry$ in such a way that $U_i^*\neq U_{i+1}^*,\,\forall\, i=0,\, 1,\, \ldots,\, q-1$. By using \eqref{eq:pont4} and \eqref{eq:defHam2} it follows that for all $t\in \lpt\lc 0,\, t_f\rpt\rp$ there exists $h_i(t)<0$ for some $i\in\ly 0,1,\ldots,n\ry$, which together with \eqref{eq:pont3} implies that
\begin{align}
h_j(t) &= \min \ly h_0(t),\,h_1(t),\,\ldots,\, h_n(t)\ry,\,\forall\, j\in \mathbb{I}_\ell,\,\forall\, t\in T_\ell.\label{eq:proofthm:hjmin}
\end{align}

In addition, from combining \eqref{eq:pont4}, \eqref{eq:defHam2}, \eqref{eq:proofthm:surfuj}, and \eqref{eq:proofthm:hjmin} it follows that the switching function corresponding to each active enzyme is given by
\begin{align}
h_i(t)&=-\frac{1}{E_T} < 0 ,\quad\forall\, i\in \mathbb{I}_\ell,\,\forall\, t \in T_\ell.\label{eq:proofseq:hi}
\end{align}
In this setup, the proof follows by showing that
\begin{align}
U_\ell^* &=\ly u_\ell\ry,\forall\, \ell=0,\,1,\,\ldots,\,q,\label{eq:proofseq:result1}\\
q&=n-1.\label{eq:proofseq:result0}
\end{align}

From \eqref{eq:defHam1} and \eqref{eq:defHam2} it can be shown that the $i$th switching function is given by
\begin{align}
h_i(t) &=  \alpha_i + (p_{i+1}(t) - p_i(t))w_i(t) ,\,\forall\,i=0,\,1,\,\ldots,\,n,\label{eq:proofthm:defhi}
\end{align}
with $p_0(t)=0$ and $p_{n+1}(t)=0$. Similarly, from \eqref{eq:pont2}, \eqref{eq:gralODE}, and \eqref{eq:defHam2} we derive the general form of the ODE for the $i$th co-state
\begin{align}
\dot{p}_i(t)&=  (p_i(t) - p_{i+1}(t))\dfrac{\partial w_i}{\partial x_i}u_i(t),\,\forall\,i=1,\,2,\,\ldots,\, n.\label{eq:defpidot}
\end{align}

We note from \eqref{eq:ass1:wnull} and \eqref{eq:finalsurf} that $x(t_f^{-})\neq 0$, which using the fact that $x(0)=0$ implies that each $u_i(t)$ must be active for some nonempty interval, i.e. for each $u_i,\, i=0,\,1,\,\ldots,\, n$ there exists an interval $R_i\neq \emptyset$ such that
\begin{align}
u_i(t)\neq 0,\,\forall\, t\in R_i.\label{eq:proofseq:Ri}
\end{align}
The proof follows using an inductive procedure based on the following result.

\begin{fact}\label{fact:indstep}
Consider interval $T_\ell,\, \ell\geq 2$, and assume that
\begin{align}
x_i(t_\ell)  &= 0,\,\forall\,i>\ell+1,\label{eq:proofseq:fac:0a}\\
U_j^* &= \ly u_j\ry,\,\forall\, j\leq \ell.\label{eq:proofseq:fac:0c}
\end{align}
Then,
\begin{align}
U_{\ell+1}^*&=\ly u_{\ell+1}\ry.\label{eq:proofseq:fac:result}
\end{align}
\end{fact}

\subsubsection*{Proof:}
From \eqref{eq:gralODE}, \eqref{eq:proofthm:defhi}, and \eqref{eq:defpidot} it can be shown that
\begin{align}
\dot{h}_j(t) &= \lp \dot{p}_{j+1}(t) - \dot{p}_j(t)\rp w_j(t) +\nonumber\\
 &\quad \lp p_{j+1}(t) - p_j(t)\rp \frac{\partial w_j}{\partial x_j}\dot{x}_j(t),\,\forall \, j <\ell,\,\forall\, t\in T_\ell,\nonumber\\
             &=   \lp p_{j+1}(t) - p_{j+2}(t)\rp\frac{\partial w_{j+1}}{\partial x_{j+1}} w_j(t) u_{j+1}(t) - \nonumber\\
         &\quad \lp p_j(t) - p_{j+1}(t)\rp\frac{\partial w_j}{\partial x_j} w_{j-1}(t)\times \nonumber\\
         &\quad u_{j-1}(t),\,\forall \, j \leq n,\,\forall\, t\in T_\ell, .\label{eq:proofseq:hjdot}
\end{align}
Equation \eqref{eq:proofseq:hjdot} shows that the only inputs that affect $\dot{h}_j(t),\,\forall\, t\in T_\ell$ are $u_{j-1}(t)$ and $u_{j+1}(t)$. Therefore, since  $u_j(t)=0,\,\forall\, j\neq \ell,\,\forall\, t\in T_\ell$ it follows that
\begin{align}
\dot{h}_j(t)&=0,\,\forall \, j <\ell-1,\,\forall\, j=\ell,\,\forall\, t\in T_\ell.\label{eq:proofseq:fac:1b}
\end{align}
On the other hand, if $j=\ell-1$ then $u_{j+1}(t)=E_T,\,\forall\, t\in T_\ell$ and $u_{j-1}(t)=0$, which after substituting in \eqref{eq:proofseq:hjdot} yields
\begin{align}
\dot{h}_{\ell-1}(t) &= \lp p_{\ell}(t) - p_{\ell+1}(t)\rp\frac{\partial w_{\ell}}{\partial x_{\ell}} w_{\ell-1}(t) E_T,\,\forall t\in T_\ell.\label{eq:proofseq:hellm1dot1}
\end{align}
Combining \eqref{eq:proofseq:hellm1dot1} and \eqref{eq:proofthm:defhi} with $i=\ell$ leads to
\begin{align}
\dot{h}_{\ell-1}(t) &= \lp \frac{\alpha_\ell - h_\ell(t)}{w_\ell(t)}\rp\frac{\partial w_{\ell}}{\partial x_{\ell}} w_{\ell-1}(t) E_T,\,\forall t\in T_\ell.\label{eq:proofseq:hellm1dot2}
\end{align}
Equation \eqref{eq:proofthm:defhi} with $i=\ell$ implies that $w_\ell(t)> 0,\,\forall\, t\in T_\ell$, since otherwise $h_\ell(t)=\alpha_\ell\geq 0$ for some $t\in T_\ell$ and \eqref{eq:proofseq:hi} cannot be satisfied. This guarantees that $\dot{h}_{\ell-1}(t)$ in \eqref{eq:proofseq:hellm1dot2} is well defined in the interval $T_\ell$. Similarly, \eqref{eq:proofthm:defhi} and \eqref{eq:proofseq:fac:0c} imply that $w_{\ell-1}(t)> 0,\,\forall\, t\in T_{\ell-1}$ and $\dot{x}_{\ell-1}(t)=0,\,\forall\, t\in T_\ell$, which means that $w_{\ell-1}(t)$ is constant in $T_\ell$ so that $w_{\ell-1}(t)>0,\,\forall\, t\in T_\ell$. Using \eqref{eq:ass1:partialw}, \eqref{eq:proofseq:hi} and $\alpha_\ell\geq 0$ in \eqref{eq:proofseq:hellm1dot2} yields
\begin{align}
\dot{h}_{\ell-1}(t)&>0,\,\forall\,t\in T_\ell.\label{eq:proofseq:fac:1c}
\end{align}

Therefore, combining \eqref{eq:proofseq:fac:0c} with \eqref{eq:proofseq:hi}, \eqref{eq:proofseq:fac:1b}, and \eqref{eq:proofseq:fac:1c} implies that the trajectory of the $j$th switching function satisfies
\begin{align}
h_j(t)&=-\frac{1}{E_T},\,\forall\, j\leq \ell,\,\forall\,t\in T_j,\label{eq:proofseq:hj}\\
\dot{h}_j(t)&>0 ,\,\forall\, j< \ell,\,\forall\,t\in T_{j+1}, \label{eq:proofseq:hjdot1}\\
\dot{h}_j(t)&=0 ,\,\forall\, j< \ell,\,\forall\, t_{j+1} < t < t_\ell.\label{eq:proofseq:hjdot2}
\end{align}

Suppose that $u_j\in U_{\ell+1}^*,\, j < \ell$, then in order to satisfy \eqref{eq:proofseq:hi}, \eqref{eq:proofseq:hj}-\eqref{eq:proofseq:hjdot2} imply that $h_j(t)$ must be discontinuous at $t=t_{\ell}$, which is not possible since both $x(t)$ and $p(t)$ are continuous. Hence, it follows that
\begin{align}
u_j &\notin U_{\ell+1}^*,\,\forall\, j < \ell. \label{eq:proofseq:fac:2a}
\end{align}

For the case when $j=\ell$, suppose that $u_\ell\in U_{\ell+1}^*$ holds, then in particular vertex $e_\ell$ is optimal $\forall\, t\in T_{\ell+1}$, which implies that $U_{\ell+1}^*=U_\ell^*$, contradicting the fact that $U_i^*\neq U_{i+1}^*,\,\forall\, i=0,\,1\,\ldots,\,q-1$, and therefore we have that
\begin{align}
u_\ell &\notin U_{\ell+1}^* .\label{eq:proofseq:fac:2b}
\end{align}

Now suppose there exists $u_j\in U_{\ell+1}^*,\, j>\ell+1$, then the linearity of $\mathcal{H}\lp x(t),p(t),u(t)\rp$ and $\mathcal{U}$ implies that in particular vertex $e_j$ is optimal $\forall\, t\in T_{\ell+1}$, which implies that
\begin{align}
\dot{x}_i(t)&=0,\,\forall\, i\neq \ly j,\,  j+1\ry.\label{eq:proofseq:xidot}
\end{align}
From \eqref{eq:proofseq:fac:0a} we have that $x_i(t_\ell)=0,\,\forall\,i>\ell+1$ and hence using \eqref{eq:proofseq:xidot} we conclude that $x_i(t)=0,\,\forall\,i>\ell+1,\, \forall\,t\in T_{\ell+1}$, but in view of \eqref{eq:proofthm:defhi} and \eqref{eq:ass1:wnull} in Assumption \ref{ass:ratelaws}B, this implies that $h_i(t)=\alpha_i\geq0,\,\forall\, i>\ell+1$, which contradicts \eqref{eq:proofseq:hi} and therefore
\begin{align}
u_j \notin U_{\ell+1}^*,\,\forall\, j > \ell+1.\label{eq:proofseq:fac:2c}
\end{align}
Therefore, \eqref{eq:proofseq:fac:result} follows from combining \eqref{eq:proofseq:fac:2a}, \eqref{eq:proofseq:fac:2b} and \eqref{eq:proofseq:fac:2c} with \eqref{eq:proofseq:Ri}.
\qed

Consider the interval $T_0$ and suppose there exists $u_j\in U_0^*,\, j>0$. Then in particular vertex $e_j$ is optimal $\forall\, t\in T_0$, which implies that $\dot{x}_i(t)=0,\,\forall i\neq j,\, i\neq j+1$. But since $x(0)=0$ it follows that $x(t)=0,\,\forall\,t\in T_0$, which from \eqref{eq:ass1:wnull} and \eqref{eq:proofthm:defhi}, implies that $h_i(t)=\alpha_i\geq0,\,\forall\, i>0$, which contradicts \eqref{eq:proofseq:hi} and therefore we conclude that
\begin{align}
U_0^*&=\ly u_0\ry.\label{eq:proofseq:2nd:1}
\end{align}

Now consider interval $T_1$ and suppose $u_0\in U_1^*$. Then in particular vertex $e_0$ is optimal $\forall\, t\in T_1$, which implies that $U_1^*=U_0^*$, contradicting our hypothesis that $U_i^*\neq U_{i+1}^*,\,\forall\, i=0,\,1\,\ldots,\,q$. Furthermore, suppose there exists $u_j\in U_1^*,\, j>1$, then in particular vertex $e_j$ is optimal $\forall\, t\in T_1$, which implies that $\dot{x}_i(t)=0,\,\forall i\neq j,\, i\neq j+1$. From \eqref{eq:proofseq:2nd:1} we have that $x_i(t_0)=0,\,\forall\,i>1$ and hence, we conclude $x_i(t)=0,\,\forall\,i>1,\, \forall\,t\in T_1$. From \eqref{eq:ass1:wnull} and \eqref{eq:proofthm:defhi}, this implies that $h_i(t)=\alpha_i\geq0,\,\forall\, i>1$, which contradicts \eqref{eq:proofseq:hi} and hence $u_j\notin U_1^*,\,\forall\, j>1$ and
\begin{align}
U_1^*&=\ly u_1 \ry.\label{eq:proofseq:2nd:2}
\end{align}
Eqs. \eqref{eq:proofseq:2nd:1} and \eqref{eq:proofseq:2nd:2} imply that $x_i(t_1)=0,\,\forall\, i>1$ and therefore, we can inductively use Fact \ref{fact:indstep}, leading to the desired result \eqref{eq:proofseq:result1}. Consider interval $T_n$, so that from \eqref{eq:proofseq:result1} it holds that $U_n^*=\ly u_n\ry$. Then from \eqref{eq:proofthm:hjmin} it follows that $h_n(t)=\min \ly h_0(t),\, h_1(t),\, \ldots,\, h_n(t)\ry,\,\forall\, t\in T_n$, but from \eqref{eq:proofseq:hj}-\eqref{eq:proofseq:hjdot2} we have that it does not exist $h_i(t),\,i\neq n$ such that $h_i(t)\leq h_n(t)$ for $t\geq t_{n-1}$, thus implying that $\dot{x}_n(t)<0,\,\forall\, t\geq t_{n-1}$. This in turn means that $\lim_{t\to \infty} x_n(t)=0$ and therefore the terminal condition \eqref{eq:finalsurf} fails and $t_f\to \infty$, which leads us to the conclusion that $u_n(t)$ is never active and \eqref{eq:proofseq:result0} follows.
\qed

\end{document}